\documentclass[12pt]{article}
\usepackage{graphicx}
\usepackage{epstopdf}
\usepackage{amssymb,amsmath,epsfig}

\begin{document}

\title{\bf Fate of Electromagnetic Field on the Cracking of PSR J1614-2230 in Quadratic Regime}

\author{M. Azam$^{1}$
\thanks{azam.math@ue.edu.pk}, S. A. Mardan$^2$ \thanks{syedalimardanazmi@yahoo.com}
and M. A. Rehman$^2$ \thanks{aziz3037@yahoo.com}\\
$^1$ Division of Science and Technology, University of Education,\\
Township Campus, Lahore-54590, Pakistan.\\
$^2$ Department of Mathematics,\\ University of the Management and Technology,\\
C-II, Johar Town, Lahore-54590, Pakistan.}

\date{}

\maketitle
\begin{abstract}
In this paper, we study the cracking of compact object PSR
J1614-2230 in quadratic regime with electromagnetic field. For this
purpose, we develop a general formalism to determine the cracking of
charged compact objects. We apply local density perturbations to
hydrostatic equilibrium equation as well as physical variables
involve in the model. We plot the force distribution function
against radius of the star with different parametric values of model
both with and without charge. It is found that PSR J1614-2230
remains stable (no cracking) corresponding to different values of
parameters when charge is zero, while it exhibit cracking (unstable)
when charge is introduced. We conclude that stability region
increases as amount of charge increases.
\end{abstract}
\section{Introduction}

Self-gravitating compact objects (CO) like neutron stars, white
dwarfs, millisecond pulsars, and so forth, belongs to a distinguish
class of those celestial bodies whose study become very significant
in novel astrophysical research. It is evident that when a star or
system of stars burns out all its nuclear fuel, its remnants can
have one of three possibilities: white dwarfs, neutron stars and
black holes. The stability of stellar remnants plays a key role in
general relativity (GR) as well as modified relativistic theories
\cite{2}. The occurrence of gravitational collapse may be as a
result of cooling of gaseous material, change in anisotropy,
fluctuation of gravitational waves and variation of electromagnetic
field of CO \cite{3}. Therefore, such phenomena stimulate our
interest to study the stability regions of these self-gravitating
CO.

Astronomical objects are not physically viable, if they are unstable
towards perturbations. Therefore, it is important to check the
stability of these objects. In this context, Bondi \cite{4}
initially developed hydrostatic equilibrium equation to examine the
stability of self-gravitating spheres. Chandrasekhar \cite{5}
calculated the principle value, i.e., $\frac{4}{3}$ to determine the
dynamical instability of sphere filled with perfect fluid in GR.
Herrera \cite{6} presented the technique of cracking to discuss
gravitational collapse of self-gravitating spherical CO. This
technique interprets the behavior of inner fluid distribution of CO
just after equilibrium state is disturbed. Cracking takes place in
CO when radial forces changes its sign from positive to negative and
vice versa \cite{7}. Several authors \cite{8}-\cite{11} studied
non-local effects of cracking through radial sound speed velocities
and Raychaudhuri equation for spherically symmetric CO. Gonzalez
\cite{new1,new2} presented the idea of local density perturbation
(DP) to discussed the idea of cracking for relativistic spheres.

To study the effect of charge on the physical properties of stars is
an important subject in GR. In this scenario, Bonnor \cite{12,13}
explored the effect of charge on spherically symmetric CO and found
that electric repulsion can halt the gravitational collapse. Bondi
\cite{14} used local Minkowski coordinates to described the
contraction of radiating isotropic spherical symmetry. The main
hindrance in astrophysics and GR is to develop stable mathematical
models which describes the characteristic of charged spherical CO.
Bekenstein \cite{15} presented the idea of gravitational collapse in
charged CO. Ray et al. \cite{16} found the maximum amount of charge,
(i.e., approximately $ 10^{20}$ coulomb), needed for CO to be in
equilibrium configuration. Some authors \cite{19,20} studied the
impact of charge on gravitational collapse of celestial objects and
analyzed the tendency of self-gravitating systems to produce charged
black holes or naked singularities. Sharif and Azam \cite{21,22}
studied the stability of spherical and cylindrical symmetric objects
under the influence of electromagnetic field.

Demorest et al. \cite{23} used the Green Bank Telescope at the
National Radio Astronomy Observatory to analyze the system of stars
by means of Shapiro delay (SD) and presented the observed values of
different physical parameters for PSR J1614-2230. These physical
parameters like ecliptic longitude, ecliptic latitude, parallax
pulsar spin, pulsar spin period, orbital period, companion mass,
radius, and so forth are recorded with very high precision by SD for
PSR J1614-2230. The availability of very accurate parametric values
made PSR J1614-2230 extremely important for modern research in GR.
Neutron stars are made of the most dense material exist in this
universe. Tauris et al. \cite{23a} developed mathematical model of
PSR J1614-2230 and provided the possible variation of masses to show
that PSR J1614-2230 was born more massive as compared to any
discovered neutron star. Lin et al. \cite{23b} used stellar
evaluation code ``MESA" to describe the relationship between PSR
J1614-2230 and its stellar companion. This discovery of high massive
neutron star has extensive consequences on the equation of state
(EoS) of matter with high densities. The relationship between
physical parameters become more complicated as linear EoS is
replaced by nonlinear EoS. In this work, we apply the concept of
cracking to self-gravitating CO in the presence of electromagnetic
field in quadratic regime. Here, we take local density perturbation
(DP) which is different from constant DP presented by Herrera
\cite{6}. We applied this technique to the model of charged compact
objects with quadratic EoS presented by Takisa et al. \cite{27} and
determine the cracking of newly discovered PSR J1614-2230 with
electromagnetic field. Recently, we have investigated the cracking
of some compact objects with and without electromagnetic field in
linear regime \cite{27a}.

This paper is arranged as follows. Section \textbf{2} deals with
Einsten-Maxwell field and Tolman–Oppenheimer–Volkoff (TOV) equations
corresponding to an- isotropic fluid. We present the general
formalism to determine the cracking of charged CO with local DP in
the quadratic regime in section \textbf{3}. Section \textbf{4}
investigate stable and unstable regions of compact star PSR
J1614-2230. In the last section, we conclude our results.

\section{Einstein-Maxwell Field and Tolman–Oppenheimer–Volkoff Equations}

We consider the line element for a static spherically symmetric
space time in curvature coordinates given by
\begin{equation}\label{1}
ds^2=-e^{2 \nu}dt^{2}
+e^{2 \lambda}dr^{2}+r^2 (d\theta^{2}+\sin^2\theta{d\phi^2}),
\end{equation}
where $0\leq{\theta}\leq{\pi},~ 0\leq{\phi}<{2\pi}$ and
$\nu=\nu(r)$, $\lambda=\lambda(r)$ are gravitational potentials. The
Maxwell's equations are defined as
\begin{equation}\label{3}
F_{a b; c}+F_{b c; a}+F_{c a; b}=0,
\end{equation}
\begin{equation}\label{4}
F^{a b}_{;b}=4 \pi J^{a},~~E_{a
b}=F_{ac}F_b^c-\frac{1}{4}g_{ab}F_{cd}F^{cd},
\end{equation}
where $F^{ab}$ is the electromagnetic field tensor, $J$ is the four
current density and $E_{ab}$ is the electromagnetic energy-momentum
tensor \cite{28}. The skew-symmetric electromagnetic field tensor
can be decomposed as
\begin{equation}\label{4b}
F^{ab}=\left[
          \begin{array}{cccc}
             0 &  E_x &  E_y &  E_z \\
          - E_x & 0   & B_z & B_y \\
          - E_y &-B_z & 0   & B_x \\
          - E_z &- B_y &-B_x & 0\\
          \end{array}
        \right],
\end{equation}
where $\textbf{E}=(E_x, E_y, E_z)$ is the electric field and
$\textbf{B}=(B_x, B_y, B_z)$ is the magnetic field. The
electromagnetic field tensor and four current density can be defined
as
\begin{equation}\label{4c}
F_{ab}=A_{b,a}-A_{a,b},~~~ J^a=\sigma u^a,
\end{equation}
where  $A$ and $\sigma$ are the four potential and proper charge
density and $u^a=e^{-\nu} \delta^a_0$ is four vector velocity of the
fluid. The four potential is defined as
\begin{equation}\label{4d}
A_a=(\phi (r),0,0,0).
\end{equation}
Using this in above equation, it yields
\begin{equation}\label{4e}
F_{01}=-\phi^{'}(r),
\end{equation}
which can also be written as
\begin{equation}\label{4f}
F^{01}=e^{-2(\nu+\lambda)} \phi'(r)=e^{-(\nu+\lambda)} E(r),
\end{equation}
where, we have used $E(r)=e^{-(\nu+\lambda)}\phi'(r)$. The total
energy-momentum tensor corresponding to charged anisotropic fluid
sphere is defined by \cite{28}
\begin{equation}\label{5}
T_{ab}=diag(-\rho -\frac{E^2}{2}, P_r-\frac{E^2}{2},
P_t+\frac{E^2}{2},P_t+\frac{E^2}{2}).
\end{equation}
The terms $E$, $\rho$, $P_r$ and $P_t$ are electromagnetic field,
energy density, radial and tangential pressure respectively.

The synergies of electromagnetic field and matter are governed by
system of field equations. These synergies of spherically symmetric
metric corresponds to Einstein-Maxwell field equations given by
\begin{equation}
G_{a b}=\kappa T_{a b}=\kappa(M_{a b}+E_{a b})
\end{equation}
where $M_{a b}$ is the energy momentum tensor for the fluid inside
the star and $E_{a b}=F_{ac}F_b^c-\frac{1}{4}g_{ab}F_{cd}F^{cd}$ is
electromagnetic field tensor. The non-zero components of
Einstein-Maxwell field equations corresponding to Eqs.(\ref{1}) and
(\ref{5}) are given as follows
\begin{eqnarray}\label{7}
 1+e^{-2\lambda}(2r\lambda^\prime-1)&=&8\pi r^2 \rho+r^2\frac{E^2}{2},
\\\label{8} 1-e^{-2 \lambda}(2r\nu^\prime+1)&=&-8\pi r^2
P_r+r^2\frac{E^2}{2}, \\\label{9} e^{-2 \lambda}(\lambda^\prime
r-\nu^\prime r - \nu^{\prime\prime} r^2+\nu^\prime \lambda^\prime
r^2-(\nu^\prime)^2r^2)&=&-8\pi r^2{P_t}-r^2\frac{E^2}{2},
\\\label{10} r^2\sigma &=&e^{- \lambda}  (r^2 E)^\prime,
\end{eqnarray}
where $``\prime"$ denotes the differentiation with respect to $r$.

It is clear that the choice of EoS of fluid inside the star plays a
key role for its physical significance. Thus, a star is physically
acceptable, if it satisfy the barotropic EoS $P_r=P_r(\rho)$. In
this work, we have used the quadratic EoS to explore the stability
of PSR J1614-2230. The quadratic EoS is given by \cite{27}
\begin{equation}\label{6}
 P_r=\gamma \rho^2+\alpha\rho-\beta,
\end{equation}
where $\gamma,~ \alpha$ and $\beta$ are constants and are
constrained by $(\rho \leq \frac{1+\alpha}{2 \gamma})$ and
$\beta=\alpha\rho_\varepsilon$, where $\rho_\varepsilon=0.5\times
10^{15} g/cm^{3}$ gives the density at the boundary surface of
sphere. It is interesting to note that this equation
reduce to linear EoS, when $\gamma=0$ \cite{27}.

Solving Eqs.(\ref{7})-(\ref{9}) simultaneously, we obtain
hydrostatic equilibrium equation (TOV) for anisotropic charged fluid
\begin{equation}\label{11}
\frac{d P_r}{dr}=\frac{2(P_t-P_r)}{r}-{(\rho + P_r )}{\nu^\prime}
+\frac{E}{8 \pi r^2} (r^2 E)^\prime,
\end{equation}
which shows that gradient of pressure is effected by charge and
anisotropy of fluid. Using the relation $e^{-2 \lambda(r)}=1-2M/r +
Q^2/r^2$ in the above equation \cite{28}, it yields
\begin{eqnarray}\label{12}
\Omega=-\frac{d P_r}{dr}+\frac{2(P_t-P_r)}{r}+ {( \rho + P_r )}
\frac{-\frac{4M}{r}+4 r^2 E^2 - 8  \pi r^2 P_r}{4r
(1-\frac{2M}{r}+r^2E^2)} + \frac{{(r^2 E)}^\prime E}{4 \pi r^2}=0,
\end{eqnarray}
where the mass function with $Q=r^2 E$ is defined as
\begin{equation}\label{13}
M =4\pi\int^{r}_0(\rho(x)+\frac{E^2}{8\pi})x^2dx.
\end{equation}

\section{Effect of Local Density Perturbation}

In this section, we perturb the equilibrium configuration of charged
CO through local DP ($\delta{\rho}$). Eq.(\ref{12}) depicts that
cracking take place in interior of spherical CO when equilibrium
state is interrupted due to change in sign of perturb force i.e.,
$\delta \Omega <0\rightarrow \delta \Omega >0$ and vice-versa. We
apply the local DP to Eq.(\ref{12}) and all the physical variables
like mass, radial and tangential pressure, electromagnetic field and
their derivatives involve in Eq.(\ref{12}), given by
\begin{eqnarray}\label{14}
P_r(\rho+\delta\rho)&=&P_r(\rho)+\frac{d P_r}{d\rho}\delta\rho,
\\\label{15} \frac {d P_r }{dr}(\rho+\delta\rho)&=&\frac {d P_r
}{dr}(\rho) +\left[\frac{d}{dr}\left(\frac{d
P_r}{d\rho}\right)+\frac{d P_r}{d\rho} \frac{d^2 \rho}{d r^2}
\frac{1}{ \frac{d \rho}{d r}}\right]\delta\rho, \\\label{16} P_t
(\rho+\delta\rho)&=&P_t (\rho)+\frac{d P_t}{d\rho}\delta\rho,
\end{eqnarray}
\begin{eqnarray}\label{17}
M(\rho+\delta\rho)&=&M (\rho)+\frac{d
M}{d\rho}\delta\rho,
\\\label{18} E(\rho+\delta\rho)&=& E(\rho)+\frac{E^\prime}{\rho^
\prime}\delta\rho, \\\label{19} E^\prime(\rho+\delta\rho)&=&E^\prime
(\rho)+\frac{E^{\prime \prime}}{\rho\prime}\delta\rho.
\end{eqnarray}
The radial sound speed ${v^2_r}$ and tangential sound speed
${v^2_t}$ are define as
\begin{equation}\label{20}
{v^2_r}=\frac{dP_r}{d\rho}~~ \& ~~ {v^2_t}=\frac{dP_t}{d\rho}.
\end{equation}
The perturb form of Eq.(\ref{12}) is given by
\begin{equation}\label{21}
\Omega = \Omega_0(\rho,P_r,~ P_r^\prime,~ P_t,~M, E, E^\prime)+\delta{\Omega},
\end{equation}
where
\begin{equation}\label{22}
\delta \Omega=\frac{\partial \Omega}{\partial\rho}\delta\rho +
\frac{\partial \Omega}{\partial P_r}\delta P_r+ \frac{\partial
\Omega}{\partial P_r^\prime}\delta P_r^\prime +\frac{\partial
\Omega}{\partial P_t}\delta P_t +\frac{\partial
\Omega}{\partial{M}}\delta{M}+\frac{\partial
\Omega}{\partial{E}}\delta{E} +\frac{\partial
\Omega}{\partial{E^\prime}}\delta{E^\prime},
\end{equation}
which can also be written as
\begin{eqnarray}\label{23}
\frac{\delta \Omega}{\delta \rho}&=&\frac{\partial
\Omega}{\partial\rho} + \frac{\partial \Omega}{\partial
P_r}{v^2_r}+\frac{\partial \Omega}{\partial
P_r^\prime}({v^2_r}^\prime+{v^2_r}\rho^{\prime\prime}({\rho}^\prime)^{-1})
+\frac{\partial \Omega}{\partial P_t}{v^2_t}^\prime\\ \notag
&+&\frac{4 \pi r^2}{{\rho}^\prime} \frac{\partial \Omega}{\partial
M} (\rho + \frac{E^2}{2}) +\frac{\partial \Omega}{\partial E}
\frac{E^{\prime}}{\rho\prime}+\frac{\partial \Omega}{\partial
E^\prime} \frac{E^{\prime\prime}}{\rho\prime}.
\end{eqnarray}
This is the fundamental equation used to determine the effects of
local DP on the cracking of charged anisotropic fluid. We will plot
the force distribution function $\frac{\delta \Omega}{\delta \rho}$
against radius $``r"$ of the star for different values of the
parameters involve in the model. Using Eq.(\ref{12}), the
derivatives involve in the above equation are given as follows
\begin{eqnarray}\label{24}
\frac{\partial \Omega}{\partial\rho}&=&\frac{-4M-16\pi r^3 P_r
+3r^3E^2}{4r^2-8 M r+4r^4E^2}, \\\label{25} \frac{\partial
\Omega}{\partial M}&=&-\frac{(\rho+P_r)(4r^2-16\pi r^4 P_r-2r^4E^2)}
{(2r^2-4 M r+2r^4E^2)^2}, \\\label{26} \frac{\partial
\Omega}{\partial P_r}&=&-\frac{2}{r}- \frac{2 M+16 \pi r^3 P_r+8\pi
r^3\rho-r^3E^2}{2r^2- 4 M r+2r^4E^2}+\frac{r^2E^2}{4r-8 M+4r^3E^2},
\end{eqnarray}
\begin{eqnarray}\label{27}
\frac{\partial \Omega}{\partial P_t}&=&\frac{2}{r},
~~~~\frac{\partial \Omega}{\partial P_r^\prime}=-1, \\\label{29}
\frac{\partial \Omega}{\partial
E}&=&-\frac{(\rho+P_r)(r^2E){(3r-10M+6r^3E^2-6\pi r^3
P_r)}}{2(r-2M+r^3E^3)^2} +\frac{2+rE^\prime}{8\pi r}, \\\label{30}
\frac{\partial \Omega}{\partial E^\prime}&=&\frac{E}{8\pi}.
\end{eqnarray}

\section{Cracking of PSR J1614-2230}

Here, we apply the formalism developed in the above section to
investigate the cracking of charged objects for the model given by
Takisa et al. \cite{27}. This model is consistent with the physical
features of observed objects and its connection can be made with PSR
J1614-2230 for particular values of parameters given in \cite{27}.
The analysis of Takisa was seems to be consistent with
observational objects such as Vela X-1, Cen X-3, SMC X-1, PSR
J1903-327 and PSR J1614-2230. But our focus in this analysis is the
particular object PSR J1614-2230 because its mass and radius has
been measured with great accuracy. The model is defined by following
equations
\begin{eqnarray}\label{31}
M(r)&=&\frac{r^3\,(4\,a-4\,b)}{8\,(a\,r^2+1)}
+\frac{5\,s\,\arctan\!\sqrt {a\,r^2}}{8\,a^{\frac{3}{2}}}\notag\\
&&-\frac{r\,s\,(-2\,a^2\,r^4+10\,a\,r^2+15)}{24\,a\,(r^2+a+ 1)},
\\\label{32}
{\rho}&=&\frac{(2\,a-2\,b)\,(a\,r^2+3)-a^2\,r^4\,s}{16\,\pi\,{(a\,r^2+1)}^2},
\end{eqnarray}
\begin{figure}
\label{111} \centering
\includegraphics[width=80mm]{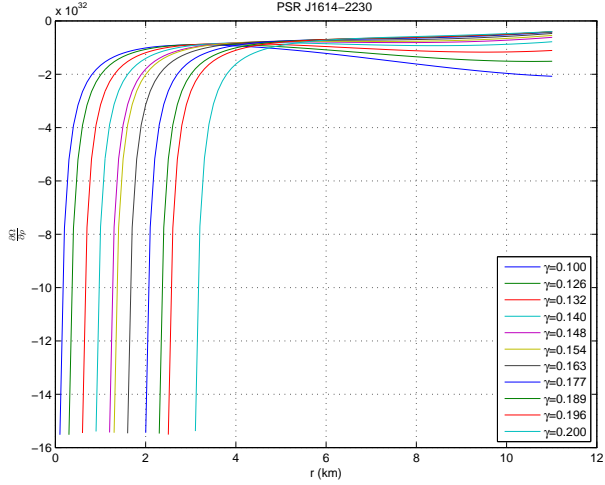}
\caption{Plots shows that there is no cracking, i.e., the PSR
J1614-2230 remains stable for different values of the parameters
involved in the model given in Table {1}, when $E=0$ in quadratic
regime.}
\end{figure}
\begin{figure}
\label{112} \centering
\includegraphics[width=80mm]{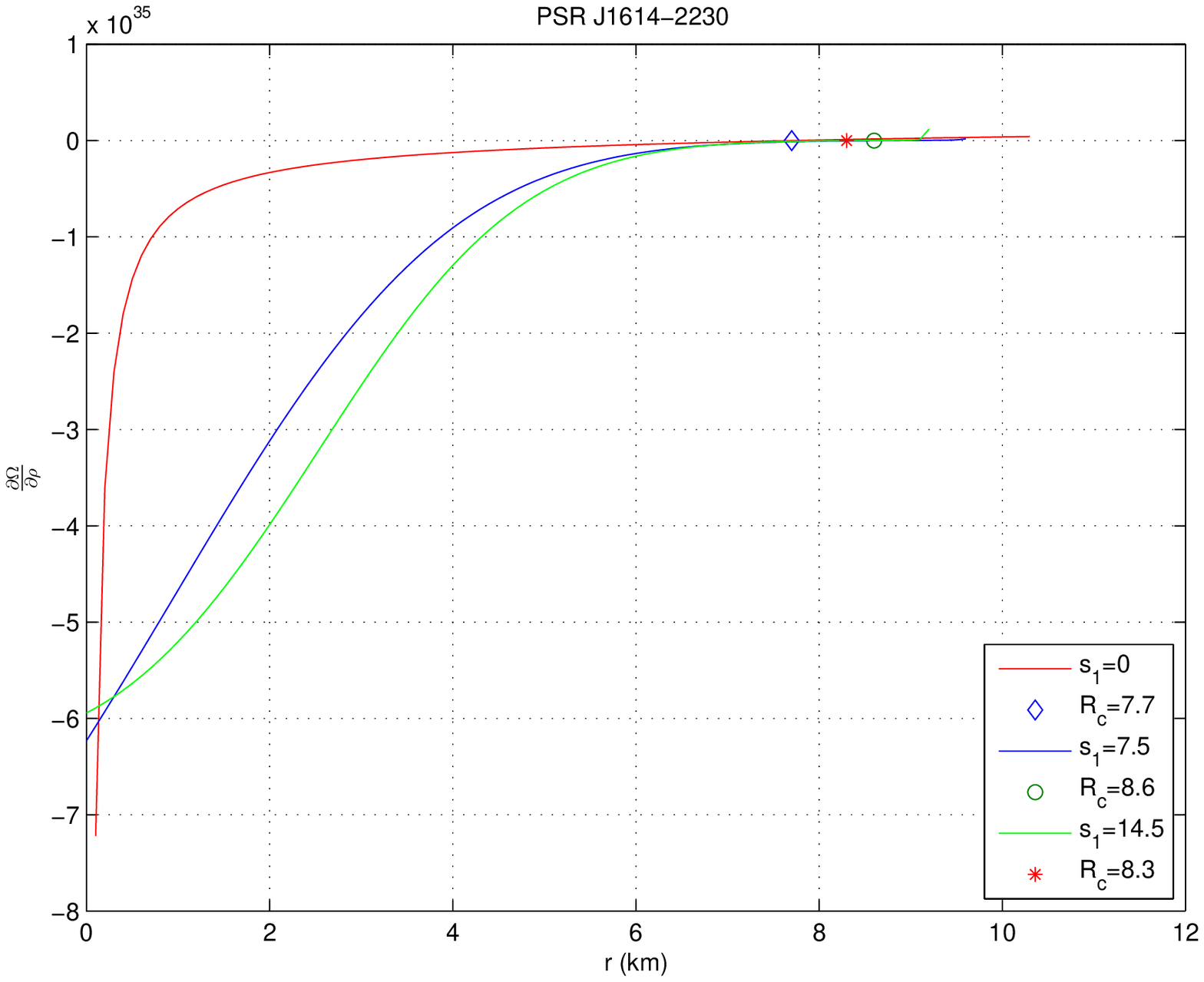}
\caption{Cracking of PSR J1614-2230 with $\gamma=0.0$, $\alpha=0.99$
and $s_1=0, 7.5, 14.5$.}
\end{figure}
\begin{eqnarray}\label{33}
P_r&=&\frac{\gamma\,{[(2\,a-2\,b)\,(a\,r^2+3)-a^2\,r^4\,s]}^2}{256\,
{\pi}^2\,{(a\,r^2+1)}^4}\notag\\
&+&\frac{\mathrm{\alpha}\,[(2\,a-2\,b)\,(a\,r^2+3)-a^2\,r^4\,s]}
{16\,\pi\,{(a\,r^2+1)}^2}-\mathrm{\beta},
\\\label{34} P_t&=&P_r+\Delta,
\end{eqnarray}
where
\begin{eqnarray}\label{35}
&&8\pi\Delta=\frac{4\,r^2\,(b\,r^2+1)}{a\,r^2+1}\,
\Bigg(\frac{\mathrm{F^{''}}}{2\, r^2}
-\frac{\mathrm{F^{'}}}{2\,r^3}+\frac{{\mathrm{F^{'}}}^2}{4\,r^2}
+\frac{b^2\,n\,(n-1)}{{(b\,r^2+1)}^2}\notag\\
&&+\frac{a^2\,t\,(t-1)}{{(a\,r^2+1)}^2}+\frac{\mathrm{F^{'}}\,
 b\,n}{r\,(b\,r^2+1)}
+\frac{\mathrm{F^{'}}\,a\,t}{r\,(a\,r^2+1)}+\frac{2\,a\,b\,n\,t}
{(a\,r^2+1)\,(b\,r^2+1)}\Bigg)\notag\\
&&\Bigg(\frac{4\,b\,r^2+4}{a\,r^2+1}-
\frac{r^2\,(2\,a-2\,b)}{{(a\,r^2+1)}^2}\Bigg)\,
\Bigg(\frac{\mathrm{F^{'}}}{2\,r}+\frac{b\,n}
{b\,r^2+1}+\frac{a\,t}{a\,r^2+1}\Bigg)\notag \\
&&-\frac{2\,a-2\,b-16\,\pi\,\mathrm{\beta}\,{(a\,r^2+1)}^2
+a^2\,r^2\,s}{2\,{(a\,r^2+1)}^2}-\frac{\mathrm{\alpha}\,((2\,a-2\,b)\,(a\,r^2+3)+a^2\,r^2\,s)}
{2\,{(a\,r^2+1)}^2}\notag \\
&&-\frac{\gamma\,((a-b)\,(a\,r^2+3)-a^2\,r^4\,s)}
{64\,\pi\,{(a\,r^2+1)}^2},
\end{eqnarray}
and
\begin{eqnarray}\label{36}
t&=&\frac{\mathrm{\alpha}}{2}+\gamma\,{(2\,a-2\,b)}^2\,
\Bigg(\frac{b}{{(a-b)}^2}-\frac{b^2}{{(a-b)}^3}
+\frac{1}{4}\Bigg)+\frac{s\,(\mathrm{\alpha}+1)}{8\,a-8\,b}\notag\\
&+&\frac{\gamma\,s\,(2\,b^3\,(2\,a-1)+(a-b)\,
(a+b+2\,s\,(a-b))-6\,a\,b^2)}{8\,{(a-b)}^3}, \\\label{37}
n&=&\frac{\mathrm{\beta}\,(a-b)}{4\,b^2}+\gamma\,
{(2\,a-2\,b)}^2\,\Bigg(\frac{b}{{(a-b)}^2}
-\frac{b^2}{{(a-b)}^3}+\frac{1}{4}\Bigg)\notag \\
&+&\frac{2\,\mathrm{\alpha}\,(a-b)}{4\,a-4\,b}+
\frac{\gamma\,s\,(2\,b\,(2\,a^3\,b
-6\,a^2\,b^2)-a^4\,(4\,b+s))}{16\,b^2\,{(a-b)}^3}\notag\\
&+&\frac{(a-b)\,(\mathrm{\alpha}+1)}{4\,b}
-\frac{a^2\,s\,(\mathrm{\alpha}+1)}{8\,b^2\,(a-b)},
\\\label{38} E^2&=&\frac{sa^2r^4}{(1+ar^2)^2}.
\end{eqnarray}
The radial and tangential sound speed velocities can be obtained
from Eqs.(\ref{33}) and (\ref{34}) as
\begin{equation}\label{39}
{v^2_r}=\mathrm{\alpha} + 2\, \gamma\, \mathrm{\rho},
\end{equation}
and
\begin{figure}
\label{113}
\centering
\includegraphics[width=80mm]{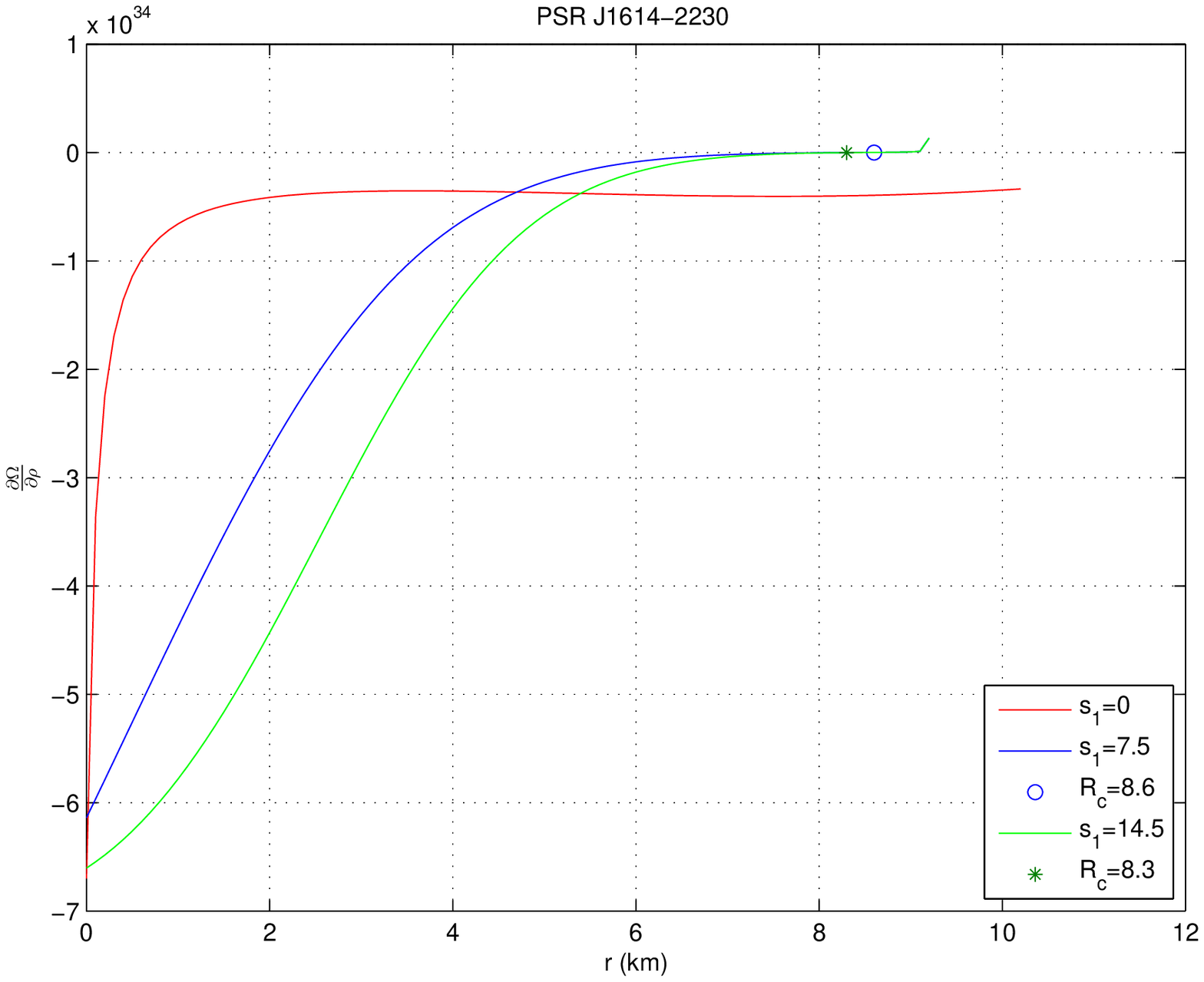}
\caption{Cracking of PSR J1614-2230 with $\gamma=0.140$,
$\alpha=0.33$ and $s_1=0, 7.5, 14.5$.}
\end{figure}
\begin{figure}
\label{114}
\centering
\includegraphics[width=80mm]{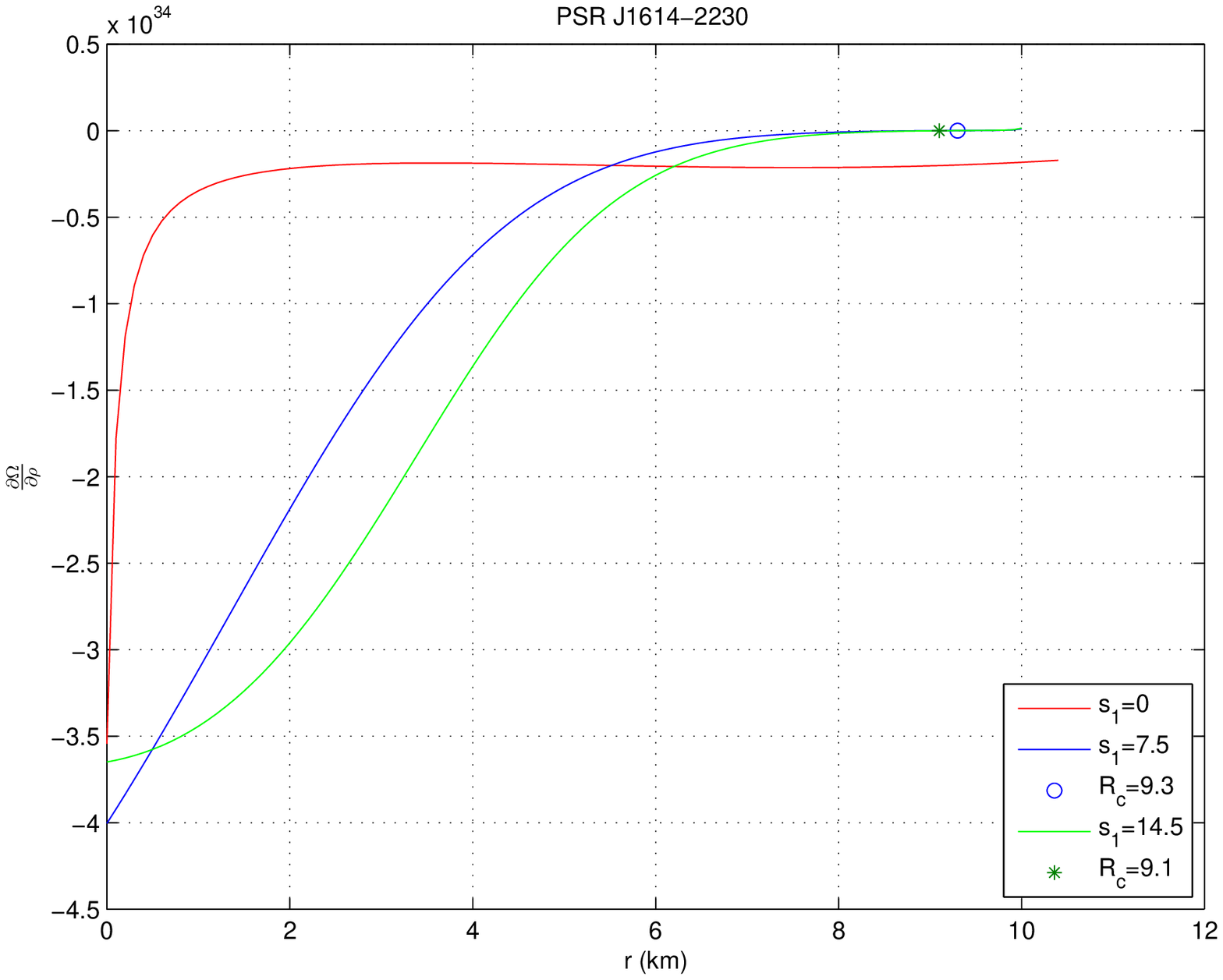}
\caption{Cracking of PSR J1614-2230 with $\gamma=0.158$,
$\alpha=0.24$ and $s_1=0, 7.5, 14.5$.}
\end{figure}
\begin{figure}
\label{115}
\centering
\includegraphics[width=80mm]{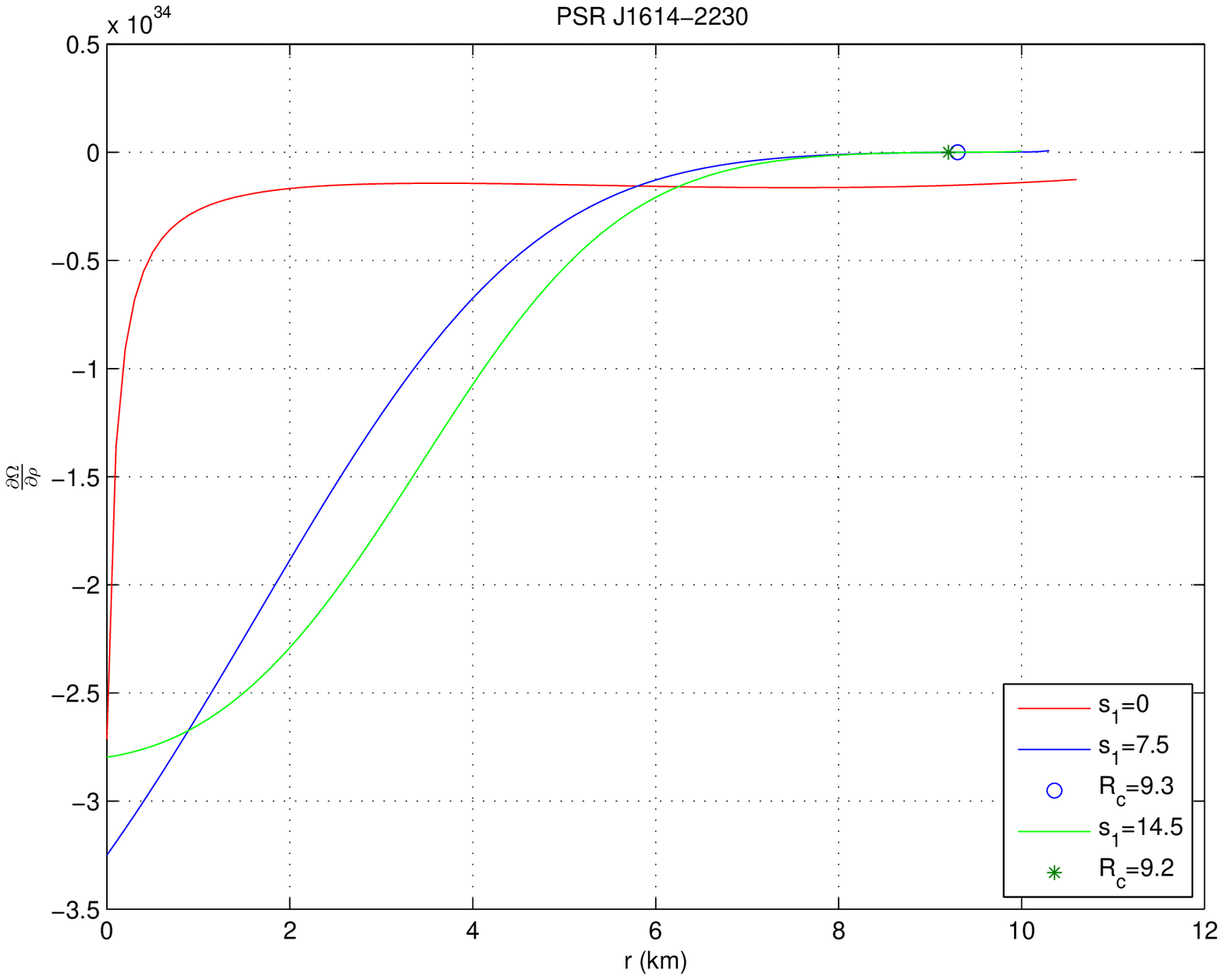}
\caption{Cracking of PSR J1614-2230 with $\gamma=0.163$,
$\alpha=0.21$ and $s_1=0, 7.5, 14.5$.}
\end{figure}
\begin{figure}
\label{116}
\centering
\includegraphics[width=80mm]{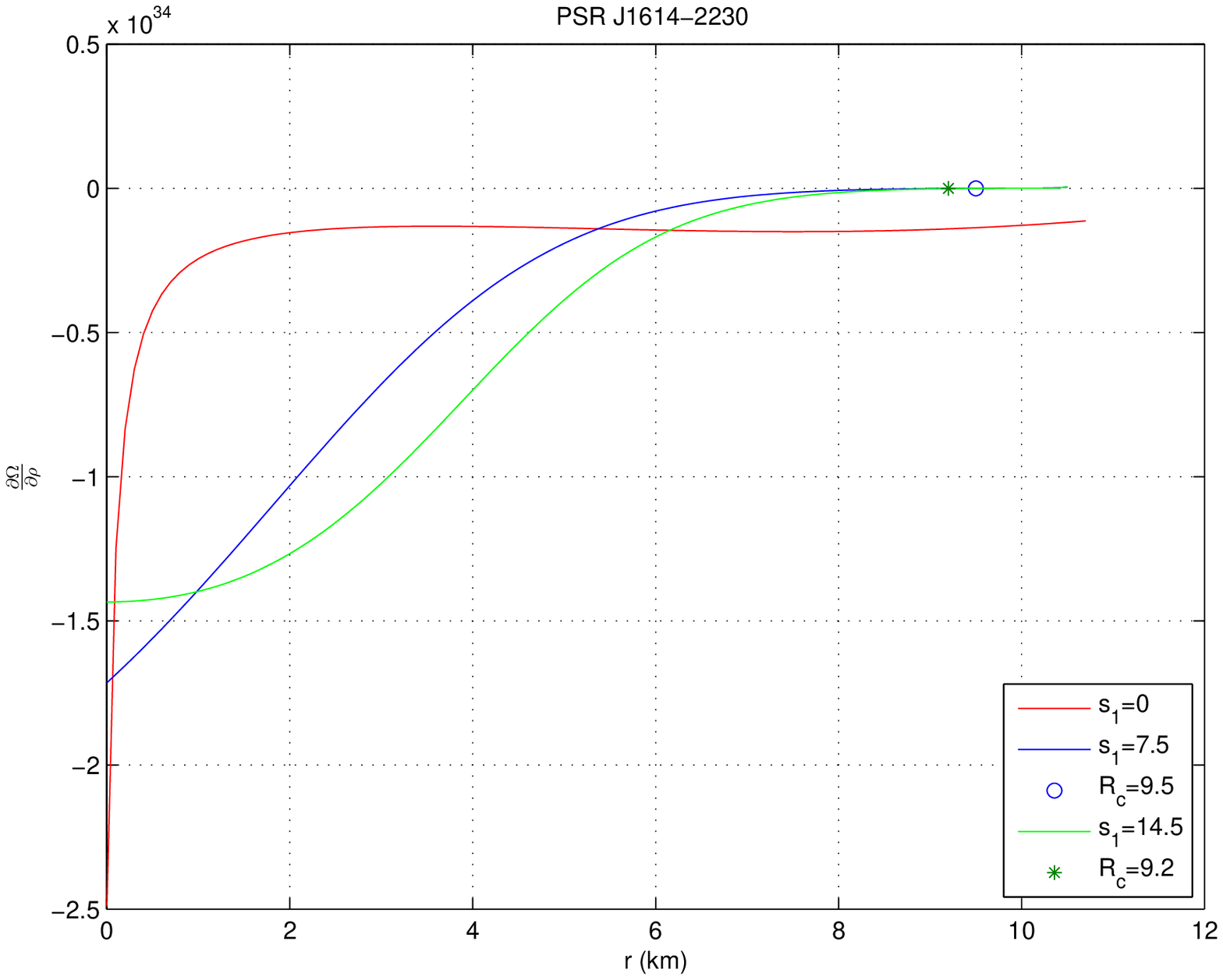}
\caption{Cracking of PSR J1614-2230 with $\gamma=0.177$,
$\alpha=0.15$ and $s_1=0, 7.5, 14.5$.}
\end{figure}
\begin{figure}
\label{117}
\centering
\includegraphics[width=80mm]{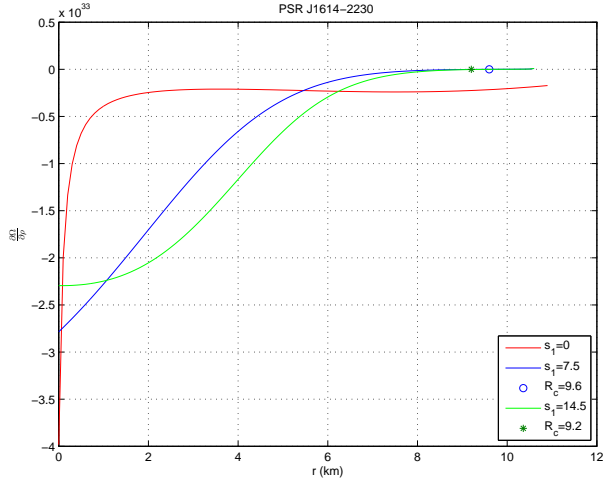}
\caption{Stability regions for $\gamma=0.196$, $\alpha=0.06$ and $s_1=0, 7.5, 14.5$}
\end{figure}
\begin{figure}
\label{118}
\centering
\includegraphics[width=80mm]{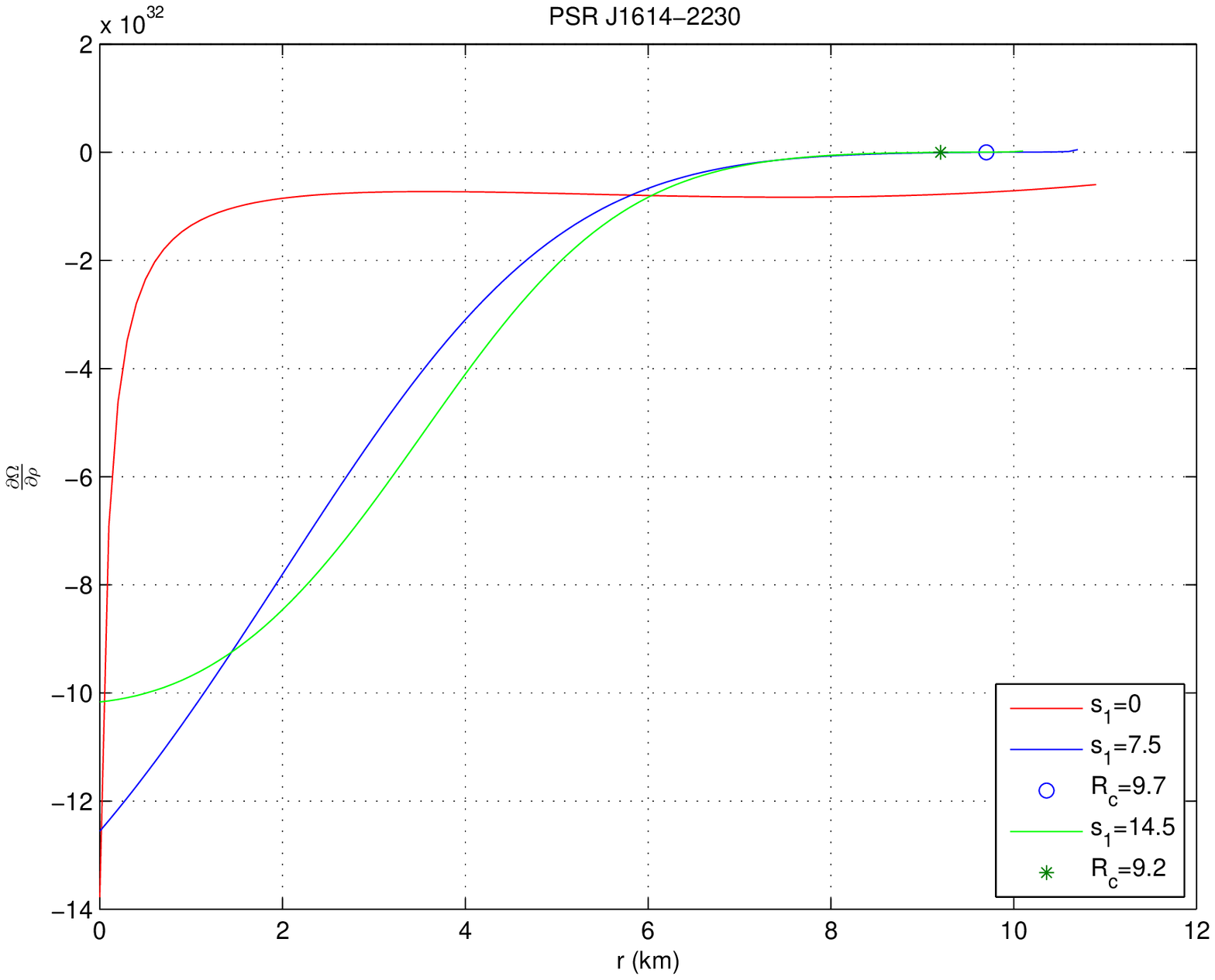}
\caption{Stability regions for $\gamma=0.200$, $\alpha=0.04$ and $s_1=0, 7.5, 14.5$}
\end{figure}
\begin{eqnarray}\nonumber
&& {v^2_t}=\Bigg[{\frac{2199023255552\gamma
((2\,a-2\,b)(3+a{r}^{2})-s{a}^{2}{r}^{4})
((4\,a-4\,b)ar-4\,s{a}^{2}{r}^{3})}
{2778046668940015(1+a{r}^{2})^{4}}}\notag\\&&
-\,{\frac{8796093022208\gamma
((2\,a-2\,b)(3+a{r}^{2})-s{a}^{2}{r}^{4})^{2}ar}
{2778046668940015(1+a{r}^{2})^{5}}}\notag\\&& +1/16\,{\frac{{\it
\alpha}\,((4\,a-4\,b)
ar-4\,s{a}^{2}{r}^{3})}{\pi\,(1+a{r}^{2})^{2}}} -1/4\,{\frac{{\it
\alpha}\,((2\,a-2\,b)(3+a{r}^{2})
-s{a}^{2}{r}^{4})ar}{\pi\,(1+a{r}^{2})^{3}}}\notag\\&&
+\Bigg\{r(1+b{r}^{2})\Bigg({\frac{t(t-1){a}^{2}}
{(1+a{r}^{2})^{2}}}+2\,{\frac {t n a b}{(1+a{r}^{2})
(1+b{r}^{2})}}+{\frac{ta{\it F^{'}}}{r(1+a{r}^{2})}}\notag\\&&
+{\frac{{b}^{2}n(n-1)}{(1+b{r}^{2})^{2}}} +{\frac{nb{\it
F^{'}}}{r(1+b{r}^{2})}} +1/2\,{\frac {{\it F^{''}}}{{r}^{2}}}
-1/2\,{\frac{{\it F^{'}}}{{r}^{3}}}+1/4\, {\frac{{{\it
F^{'}}}^{2}}{{r}^{2}}}\Bigg) (1+a{r}^{2})^{-1} \notag\nonumber
\end{eqnarray}
\begin{eqnarray}
\notag\\&& +{r}^{3}b\Bigg({\frac{t(t-1){a}^{2}}{(1+a{r}^{2})^{2}}}
+2\,{\frac{t n a b}{(1+a{r}^{2})(1+b{r}^{2})}} +{\frac{ta{\it
F^{'}}}{r(1+a{r}^{2})}}\notag\\&&+ {\frac
{{b}^{2}n(n-1)}{(1+b{r}^{2})^{2}}} +{\frac{nb{\it
F^{'}}}{r(1+b{r}^{2})}}+1/2\, {\frac{{\it
F^{''}}}{{r}^{2}}}-1/2\,{\frac{{\it F^{'}}}{{r}^{3}}}
+1/4\,{\frac{{{\it F^{'}}}^{2}}{{r}^{2}}}\Bigg)(1+a{r}^{2})^{-1}
\notag\\&&-{r}^{3}(1+b{r}^{2})
\Bigg({\frac{t(t-1){a}^{2}}{(1+a{r}^{2})^{2}}} +2\,{\frac{t n a
b}{(1+a{r}^{2})(1+b{r}^{2})}} +{\frac{ta{\it
F^{'}}}{r(1+a{r}^{2})}}\notag\\&&
+{\frac{{b}^{2}n(n-1)}{(1+b{r}^{2})^{2}}} +{\frac{nb{\it
F^{'}}}{r(1+b{r}^{2})}}+1/2\, {\frac{{\it
F^{''}}}{{r}^{2}}}-1/2\,{\frac{{\it F^{'}}}{{r}^{3}}}
+1/4\,{\frac{{{\it F^{'}}}^{2}}{{r}^{2}}}\Bigg)\notag \\
&&a(1+a{r}^{2})^{-2}+1/2\,{r}^{2}(1+b{r}^{2})
\Bigg(-4\,{\frac{t(t-1){a}^{3}r}{(1+a{r}^{2})^{3}}}
-4\,{\frac {tn{a}^{2}b r}{(1+a{r}^{2})^{2}(1+b{r}^{2})}}
\notag\\&&-4\,{\frac{t n a{b}^{2}r}{(1+a{r}^{2})
(1+b{r}^{2})^{2}}}-{\frac{ta{\it F^{'}}}{{r}^{2}(1+a{r}^{2})}}
-2\,{\frac{t{a}^{2}{\it F^{'}}}{(1+a{r}^{2})^{2}}}-4\,
{\frac{{b}^{3}n(n-1)r}{(1+b{r}^{2})^{3}}}\notag\\&&
-{\frac{nb{\it F^{'}}}{{r}^{2}(1+b{r}^{2})}}-2\,
{\frac{{b}^{2}n{\it F^{'}}}{(1+b{r}^{2})^{2}}}
-{\frac{{\it F^{''}}}{{r}^{3}}}+3/2\,{\frac{{\it F^{'}}}{{r}^{4}}}
-1/2\,{\frac{{{\it F^{'}}}^{2}}{{r}^{3}}}\Bigg)(1+a{r}^{2})^{-1}\notag\\
&&+1/8\,\Bigg({\frac{(-4\,a+4\,b)r}{(1+a{r}^{2})^{2}}}
-{\frac{(-8\,a+8\,b){r}^{3}a}{(1+a{r}^{2})^{3}}} +8\,{\frac{b
r}{1+a{r}^{2}}}\notag\\&&-{\frac{(8+8\,b{r}^{2})ar}
{(1+a{r}^{2})^{2}}}\Bigg)\Bigg({\frac{ta}{1+a{r}^{2}}}+{\frac
{nb}{1+b{r}^{2}}}+1/2\,{\frac{{\it F^{'}}}{r}}\Bigg)
\notag\\
&&+1/8\,\Bigg({\frac{(-2\,a+2\,b){r}^{2}}{(1+a{r}^{2})
^{2}}}+{\frac{4+4\,b{r}^{2}}{1+a{r}^{2}}}\Bigg)
\Bigg(-2\,{\frac{t{a}^{2}r}{(1+a{r}^{2})^{2}}}-2\,
{\frac{{b}^{2}nr}{(1+b{r}^{2})^{2}}}-1/2\,{\frac{{\it
F^{'}}}{{r}^{2}}}\Bigg)\notag\\&&-{\frac{1}{512}}\,{\frac
{\gamma((2\,a-2\,b)ar-4\,s{a}^{2}{r}^{3})}{\pi\,(
1+a{r}^{2})^{2}}}+{\frac{1}{128}}\,{\frac{\gamma((a-b)(
3+a{r}^{2})-s{a}^{2}{r}^{4})ar}{\pi\,(1+a{r}^{2})^{3}}} \notag\\
&&+1/8\,{\frac{(4\,a-4\,b+2\,s{a}^{2}{r}^{2}-32\,\pi
\,{\it\beta}\,(1+a{r}^{2})^{2})ar}{(1+a{r}^{2}) ^{3}}}\notag \\
&&-1/16\,{\frac{{\it\alpha}\,((4\,a-4\,b)
ar+2\,s{a}^{2}r)}{(1+a{r}^{2})^{2}}}+1/4\,{\frac {{\it
\alpha}\,((2\,a-2\,b)(3+a{r}^{2})+s{a}^{2}{r}^{2})ar}{(
1+a{r}^{2})^{3}}}\Bigg\}{\pi}^{-1}\Bigg]\notag\\&&
\Bigg({\frac{(1/4\,a-1/4\,b)ar-1/4\,s{a}^{2}{r}^{3}}{\pi\,(
1+a{r}^{2})^{2}}}-{\frac{((1/2\,a-1/2\,b)(3+a{r}^{2})
-1/4\,s{a}^{2}{r}^{4})ar}{\pi\,(1+a{r}^{2})^{3}}}\Bigg) ^{-1}
\notag\nonumber
\end{eqnarray}
\begin{eqnarray}
\notag\\&&-1/8\,{\frac {s{a}^{2}r-32\,\pi\,{\it \beta}\,
(1+a{r}^{2})ar}{(1+a{r}^{2})^{2}}}.
\end{eqnarray}
\begin{table}[h] \caption{Stability of neutral PSR J1614-2230 when $\gamma$ and $b_1$ are variable }
\centering
\vspace{0.5cm}
\begin{tabular}{|c|c|c|c|c|c|c|}
  \hline
$\gamma$ & $a_1$   & $b_1$   & $\alpha$& $r$(km)& $R_c$(km)\\    \hline
$0.100$  & $53.34$ & $ 6.90$ & $0.33$   & $11.07$& Stable\\\hline
$0.126$  & $53.34$ & $ 8.74$ & $0.33$   & $10.85$& Stable\\\hline
$0.132$  & $53.34$ & $10.74$ & $0.33$   & $10.60$& Stable\\\hline
$0.140$  & $53.34$ & $13.33$ & $0.33$   & $10.30$& Stable\\\hline
$0.148$  & $53.34$ & $15.61$ & $0.33$   & $9.99$ & Stable\\\hline
$0.154$  & $53.34$ & $16.87$ & $0.33$   & $9.82$ & Stable\\\hline
$0.163$  & $53.34$ & $19.04$ & $0.33$   & $9.51$ & Stable\\\hline
$0.177$  & $53.34$ & $21.72$ & $0.33$   & $9.13$ & Stable\\\hline
$0.189$  & $53.34$ & $23.64$ & $0.33$   & $8.83$ & Stable\\\hline
$0.196$  & $53.34$ & $24.73$ & $0.33$   & $8.65$ & Stable\\\hline
$0.200$  & $53.34$ & $28.42$ & $0.33$   & $8.04$ & Stable\\\hline
\end{tabular}\label{t1}
\end{table}
\begin{table}[h] \caption{Stability of PSR J1614-2230 when $s_1=0$ and $\alpha,~\gamma$ are variable}
\centering \vspace{0.5cm}
\begin{tabular}{|c|c|c|c|c|c|c|}
  \hline
$\gamma$ & $a_1$   & $b_1$   & $\alpha$ & $r$(km) & $R_c$(km)\\    \hline
$0.0  $  & $53.34$ & $13.33$ & $0.99$   & $10.30$ & 7.7\\\hline
$0.140$  & $53.34$ & $13.33$ & $0.33$   & $10.30$ & Stable\\\hline
$0.158$  & $53.34$ & $13.33$ & $0.24$   & $10.50$ & Stable\\\hline
$0.163$  & $53.34$ & $13.33$ & $0.21$   & $10.70$ & Stable\\\hline
$0.177$  & $53.34$ & $13.33$ & $0.15$   & $10.90$ & Stable\\\hline
$0.196$  & $53.34$ & $13.33$ & $0.06$   & $11.06$ & Stable\\\hline
$0.200$  & $53.34$ & $13.33$ & $0.04$   & $11.09$ & Stable\\\hline
\end{tabular}\label{t2}
\end{table}
\begin{table}[h] \caption{Stability of PSR J1614-2230 when $s_1=7.5$ and $\alpha,~\gamma$ are variable}
\centering \vspace{0.5cm}
\begin{tabular}{|c|c|c|c|c|c|c|}
  \hline
$\gamma$ & $a_1$   & $b_1$   & $\alpha$ & $r$(km) & $R_c$(km)\\    \hline
$0.0  $  & $53.34$ & $13.33$ & $0.99$   & $9.67 $ & $8.6$ \\\hline
$0.140$  & $53.34$ & $13.33$ & $0.33$   & $9.67 $ & $8.6$ \\\hline
$0.158$  & $53.34$ & $13.33$ & $0.24$   & $10.07$ & $9.3$ \\\hline
$0.163$  & $53.34$ & $13.33$ & $0.21$   & $10.37$ & $9.3$ \\\hline
$0.177$  & $53.34$ & $13.33$ & $0.15$   & $10.56$ & $9.5$ \\\hline
$0.196$  & $53.34$ & $13.33$ & $0.06$   & $10.65$ & $9.6$ \\\hline
$0.200$  & $53.34$ & $13.33$ & $0.04$   & $10.64$ & $9.7$ \\\hline
\end{tabular}\label{t3}
\end{table}
\begin{table}[h] \caption{Stability of PSR J1614-2230 when $s_1=14.5$ and $\alpha,~\gamma$ are variable}
\centering \vspace{0.5cm}
\begin{tabular}{|c|c|c|c|c|c|c|}
  \hline
$\gamma$ & $a_1$     & $b_1$    & $\alpha$ & $r$(km) & $R_c$(km)\\    \hline
$0.0  $  & $53.34$ & $13.33$ & $0.99$   & $9.21 $ & $8.3$ \\\hline
$0.140$  & $53.34$ & $13.33$ & $0.33$   & $9.21 $ & $8.3$ \\\hline
$0.158$  & $53.34$ & $13.33$ & $0.24$   & $10.05$ & $9.1$ \\\hline
$0.163$  & $53.34$ & $13.33$ & $0.21$   & $10.10$ & $9.1$ \\\hline
$0.177$  & $53.34$ & $13.33$ & $0.15$   & $10.15$ & $9.2$ \\\hline
$0.196$  & $53.34$ & $13.33$ & $0.06$   & $10.18$ & $9.2$ \\\hline
$0.200$  & $53.34$ & $13.33$ & $0.04$   & $10.19$ & $9.2$ \\\hline
\end{tabular}\label{t4}
\end{table}

The constants $a,~b$ and $s$ have dimension of length $(L^{-2})$ and
chosen in such a way that the given system satisfy the following
conditions
\begin{itemize}
  \item energy density must remains positive before and after equilibrium state.
  \item radial pressure should be vanishes at the boundary of star.
  \item At the center of star i.e. $r=0$, we have $ P_r = P_t= \Delta =0$.
  \item $v_r^2$ is constant in the quadratic regime.
  \item Across boundary of star, when $ r = \varepsilon$, we have\\
   $e^{-2 \lambda}=1-2M/\varepsilon + Q^2/\varepsilon^2$\\
   $e^{-2 \nu}=1-2M/\varepsilon + Q^2/\varepsilon^2$
\end{itemize}
By considering above conditions, we have
\begin{equation*}
\alpha=0.33, ~~a=\frac{a_1}{\Re^2},~~ b=\frac{b_1}{\Re^2},~~ s=\frac{s_1}{\Re^2}
\end{equation*}
where $\Re$=43.245$~km$ and the values given above are compatible
with observational values given by Takisa et al. \cite{27}.

For the sake of regions (stable and unstable) of PSR J1614-2230, we
have plotted force distribution function against radius of the star
for different values of the parameters involve in the model shown in
Figures \textbf{1-8}. We summarizes these results as follows:
\begin{itemize}
\item{Figure \textbf{1} depicts that all curves do not change
its sign with different values of $\gamma$ and $b_1$ corresponding
to Table \textbf{1}. Hence, we find that PSR J1614-2230 is stable in
the absence of charge in quadratic regime and it is unstable in
linear regime which is analogous to the results found in \cite{27a}.
From Table \textbf{2}, it is clear that any variation in
coefficients of quadratic EoS does not effect stability even radius
of PSR J1614-2230 changes approximately to $4\%$.}
\item{In Figure \textbf{2}, there are three curves corresponding to model parameters
$\gamma=0.0,~\alpha=0.99$ and charge $s_1=0,~7.5,~14.5$. It is noted
that all three curves (red, blue and green) change its sign for
charge parameter ($s_1=0,~7.5,~14.5$), respectively in the linear
regime. This shows that PSR J1614-2230 is unstable in linear regime,
where the symbols $``\diamond",~``o",~``*"$ represents the cracking
points (where curve changes its sign from negative to positive)
corresponding to $s_1=0,~7.5,~14.5$, respectively. The cracking
values $R_c=7.7,~8.6,~8.3$ corresponds to $s_1=0,~7.5,~14.5$ (red,
blue and green) are given in (Table \textbf{2,3,4}). In this case,
our results are consisted with \cite{27a} in linear regime.}
\item{Figures
\textbf{3-8} represents the cracking of PSR J1614-2230 star for
fixed values of parameters $\gamma=0.140, 0.158, 0.163, 0.177,
0.196, 0.200$, $\alpha=0.33,0.24,0.21, 0.15, 0.06, 0.04$ and charge
$s_1=0.0,7.5,14.5$ in quadratic regime. We see that cracking take
place for charge parameter $s_1=7.5$ and $14.5$, which are indicated
by the cracking points $``o"$ and $``*"$ corresponding to blue and
green curves, respectively. These cracking points ($R_c$) are given
in Table \textbf{3-4}. However, in each case, the star remain
stable, i.e., no cracking take place for $s_1=0$ in quadratic
regime. Hence, PSR J1614-2230 exhibit cracking both in linear and
quadratic regime when charge is present. From these illustrations,
we conclude that as charge increases cracking points are slightly
shifted towards center, which indicates that binding forces of CO
become stronger and more mass is directed towards origin.}
\end{itemize}

\section{Conclusions and observations}

We have applied the technique of cracking presented by Herrera
\cite{6} to charged anisotropic self-gravitating CO. The impact of
local DP on the stability of inner fluid of star in the presence of
charge is considered in the scenario of GR. It has been observed
that cracking of CO takes place when the system leave its
equilibrium state. The numerical value of $R_c$ (cracking point)
provides the stable/unstable region in the quadratic regime.

We have used the model of Takisa et al. \cite{27} to investigate the
cracking of PSR J1614-2230 with and without charge. Figure
\textbf{1} represents the stability of celestial object PSR
J1614-2230 in the absence of charge for different values of
parameters $\gamma$ and $b_1$ given in Table \textbf{1}. It is shown
that PSR J1614-2230 remains stable, when quadratic EoS is considered
in neutral case but it exhibited cracking with the inclusion of
charge. Figure \textbf{2} has been plotted for $\gamma=0.0$ and
different values of $\alpha$ and charge $s_1=0,~7.5,~14.5$. It is
shown that PSR J1614-2230 exhibit cracking in each case given by
$R_c=7.7,8.6,8.3$. For $s_1=0$, PSR J1614-2230 shows cracking which
is consistent with our recent published work \cite{27a}. It is worth
mentioned here that our results are analogous to \cite{27a}, when
$\gamma=0$ (Linear Regime) in the presence of charge.

In figures \textbf{3-8}, we have given the comparison of stability
region with different values of $\alpha,~\gamma$ and charge
parameter $s_1$. In these figures stability regions are plotted for
values of charge parameter $s_1=0, 7.5, 14.5$ with $R_c$ represented
by $``\diamond"$, $``o"$ and $``*"$ for $s_1=0.0$, $s_1=7.5$ and
$s_1=14.5$, respectively. We observe that the value of $R_c$
decreases as electromagnetic field increases, which are given in
Tables \textbf{2-4} for different values of parameters. Figures
\textbf{3-8} shows that cracking take place in each case for
different values of the parameters corresponding to $s_1=7.5$ and
$s_1=14.5$ in the quadratic regime, while remains stable in the
absence of charge ($s_1=0.0$).

It is noted that the local DP scheme does not affect the stability
of CO (remains stable) in neutral case, while change its stability
(potentially unstable) drastically with the inclusion of charge in
quadratic regime. Thus, the local DP scheme under nonlinear EoS
considerably effect the stability regions of CO. When physical
parameters like mass, electromagnetic field and density of
anisotropic charged self-gravitating CO are locally perturbed, they
drastically affect the sensitivity of radial forces which may leads
towards the gravitational collapse. Therefore, the stability region
of PSR J1614-2230 increases as value of electromagnetic field
increases. Hence, we conclude that the binding forces of neutron
star PSR J1614-2230 become stronger as we move towards center of
star and it becomes more dense as charge increases.

It is important to mentioned here that the idea of cracking was
presented by Herrera \cite{6} to understand the behavior of inner
fluid distribution just after departure from equilibrium state may
be responsible for cracking (overturning) of anisotropic sphere
\cite{9}. In his study, the global DP affects physical quantities
like mass, tangential and radial pressure but do not effect pressure
gradient. In this work global DP technique is modified by local
density perturbations to study cracking in the presence of
electromagnetic field. Finally, we conclude that the the given
object exhibits cracking in the presence of electromagnetic field.

\end{document}